\PassOptionsToPackage{table}{xcolor} 
\documentclass[twocolumn,trackchanges]{aastex701}

\usepackage{amsmath}
\usepackage{siunitx}
\usepackage{siunitx}
\usepackage{rotating} 
\makeatletter
\@ifpackageloaded{multirow}{}{%
  \usepackage{multirow}
}
\makeatother

\begin{document}

\title{Wind Accretion in Massive Binaries Experiencing High Mass Loss Rates: II. Eccentricity}

\shorttitle{}
\shortauthors{B. Mukhija and A. Kashi}

\author[0009-0007-1450-6490]{Bhawna Mukhija}
\affiliation{Department of Physics, Ariel University, Ariel, 4070000, Israel}
\email{bhawnam@ariel.ac.il}

\author[0000-0002-7840-0181]{Amit Kashi}
\email{kashi@ariel.ac.il}
\affiliation{Department of Physics, Ariel University, Ariel, 4070000, Israel}
\affiliation{Astrophysics, Geophysics, and Space Science (AGASS) Center, Ariel University, Ariel, 4070000, Israel}

\begin{abstract}

We perform numerical simulations to investigate high-power wind accretion in massive binary systems undergoing enhanced mass-loss episodes. The primary star is taken in the mass range $M_{1} = 60$--$90\,\mathrm{M_{\odot}}$, while the companion is a $30\,\mathrm{M_{\odot}}$ hot star. We model binary orbits with eccentricities of $e = 0$--$0.6$ and orbital periods of $P=455$--$1155$ days. We initiate strong eruptive events for the primary with mass-loss rates of $\dot{M}_{\rm w} = 10^{-2}$ -- $10^{-1}\,\rm{M_{\odot}~{yr}^{-1}}$, lasting for $1.5$ years. A fraction of the ejected wind material is accreted by the companion, with the accretion efficiency determined by the orbital separation, eccentricity, and stellar mass ratio. We analyze the resulting accretion rates and provide an analytical relation describing their dependence on the stellar mass ratio, mass-loss rate, and orbital parameters. We find that although the accretion modifies the stellar parameters of the secondary, the companion remains in thermal equilibrium and does not undergo significant radial expansion.
We further include wind mass loss from the companion during wind accretion and find a substantial reduction in accretion efficiency compared to no wind scenario. For longer orbital periods, the models yield negative accretion rates, implying that any captured material is expelled or prevented from settling onto the accretor.
These results provide new insight into the role of eccentric orbits and extreme mass-loss events in shaping the mass-transfer processes in massive binaries.
\end{abstract}

\

    
\section{Introduction}

Mass transfer (MT) via stellar winds is a fundamental process in massive binaries \citep[e.g.,][]{1992ApJ...391..246P, 1998A&ARv...9...63V, 2003ApJ...597..513S,2004MNRAS.350.1366S, 2010ApJ...709L..11K, 2014ApJS..215...15S, 2017ApJS..230...15M,2024ARA&A..62...21M, Mukhija_2025, 2026NewA..12202475M, 2025RAA....25b5010B}, which occurs whenever the primary’s outflow can be gravitationally focused and partially captured by its companion in the system. The classical framework is Bondi--Hoyle--Lyttleton (BHL) accretion, in which a gravitating object moving supersonically through a gas focuses material within an accretion radius and accretes at a rate that scales with the ambient density and the relative velocity \citep{1944MNRAS.104..273B,2004NewAR..48..843E}. In binaries, deviations from the ideal BHL assumptions are common because the relative speed combines the wind’s terminal velocity ($v_{\mathrm{wind}} $) with the orbital velocity ($v_{\mathrm{orb}} $), while spatially varying density and velocity fields introduce strong geometry and phase dependence \citep{2004NewAR..48..843E}. The winds of hot, luminous massive stars are radiatively driven by line scattering \citep{Lucy2007, Vink2001, 2004ApJ...614..929C, 2007ASPC..361..153K, 2025A&A...703A.279M, 2025arXiv251213782S}. Their mass-loss rates and terminal velocities depend on stellar parameters and metallicity, exhibiting features such as the bi-stability jump and showing intrinsic time-dependent structure due to the line-driven instability \citep{Owocki1994, Lucy2007, Vink2001}. 

For massive primary stars in binaries, these properties set the density and velocity profiles that control wind capture and thus the accretion rate onto the companion.
In many cases BHL provides a useful approximation for the accretion rate \citep[e.g.,][]{2007MNRAS.378.1609K}. In other cases, multidimensional simulations and analytic arguments show that the effective accretion efficiency can differ from BHL predictions due to various effects, such as orbital deflection, shocks, angular-momentum transport, finite size of the accretor \citep[e.g.,][]{Nagae2004, ElMellah2015, MacLeod2015,2020MNRAS.492.5261K}.
Moreover, when the accretor has its own wind, the accretion rate can become sub-BHL due to the opposing wind momentum \citep{2022MNRAS.516.3193K}.
Eccentric orbits further modulate the instantaneous accretion rate by varying the separation and relative velocity across the orbit, often leading to periastron-enhanced episodes of capture \citep{2010MNRAS.405.1924K,2017MNRAS.464..775K}.

\begin{center}
\begin{table*}
        \begin{tabular}{|l|c c c c c c| c|}
    \hline
    \hline
    & \multicolumn{6}{c}{Primary stars} & Companion star \\  
    \hline
    
    Stellar parameter &  60$~\rm M_{\odot}$ & 70$~\rm M_{\odot}$ & 80$~\rm M_{\odot}$ & 90$~\rm M_{\odot}$ & 100$~\rm M_{\odot}$ &  &30$\rm~ M_{\odot}$\\
    \hline
     $  M ~( \rm M_{\odot}$) & 49.37 & 54.44 & 58.45 & 61.67 & 63.29 & & 29.07   \\ 
     $\log T_{\rm eff}~ (\rm K)$ & 4.50 & 4.42 & 4.40 & 4.36  & 4.28 &  & 4.57  \\
     $\log R ~(\rm R_{\odot}) $ & 1.46 & 1.67 & 1.78 & 1.89  & 2.07 & & 0.98  \\
     $\log g ~(\rm cm~s^{-2})$ & 3.22 & 2.82 & 2.63 & 2.43& 2.08 &  & 3.93  \\
     $\log L ~(\rm L_{\odot})$ & 5.86 & 6.00 & 6.10 & 6.19  & 6.25 &  &  5.20  \\
     $\log \dot{M}~(\rm {M_{\odot}}\rm~yr^{-1})$ &-5.43 & -5.39 & -5.11 &-4.72   & -4.12 & & -6.39  \\
    \hline 
    \hline
    \end{tabular}
    \caption{Stellar parameters correspond to the $ \rm 60$--$100~\rm{M_{\odot}}$ primary stars, and $ \rm 30~M_{\odot}$ companion star and at the initial stage of wind accretion. The rows from first to last represent: mass of the star ($ M$) at the time of wind accretion, surface temperature ($ T_{\rm eff}$), surface gravity ($ g $), surface luminosity ($ L$), and mass loss via stellar winds ($ \dot{M}$), respectively (see Figure~\ref{fig_5}).}
    \label{T1}

\end{table*}
\end{center}

\begin{figure*}
    \centering
    \includegraphics[width=1\linewidth]{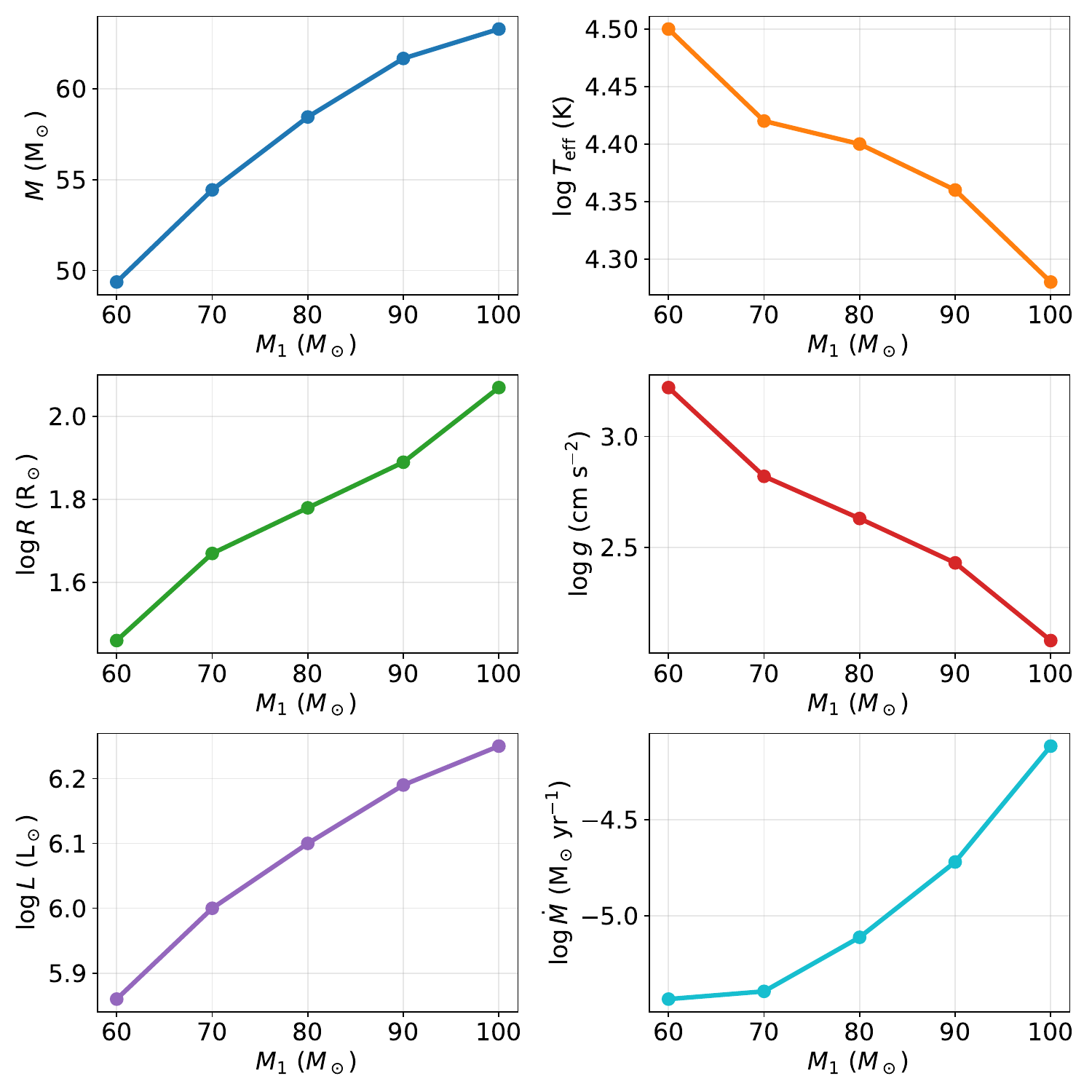}
    \caption{Multi-panel visualisation of the stellar parameters listed in Table~\ref{T1} as a function of primary mass $M_1$ at the onset of wind accretion. The panels show stellar mass, effective temperature, radius, surface gravity, luminosity, and wind mass-loss rate. The figure highlights the systematic trends of the primary star properties across the $60$–$100~\rm M_\odot$ models.
}
    \label{fig_5}
\end{figure*}

In our previous work \citep{Mukhija_2025}, we examine BHL wind accretion in massive binary systems using \textsc{mesa} simulations, focusing on how accretion rates depend on stellar and orbital properties. We modeled systems with mass ratios \( q = 0.3 \)-0.8 (companion masses \( M_{2} = 30-80~\rm {M}_{\odot} \)) while fixing the primary at \( M_{1} = 100~\rm {M}_{\odot} \). Each model was evolved for 1.5~years with a constant primary mass-loss rate. Our results showed that accretion rates increase with higher mass ratios, reflecting the stronger gravitational focusing of more massive companions. A mild upward trend in accretion rate over time was seen in wider systems, likely caused by reduced relative wind--companion velocities at longer orbital periods, which enhance transfer efficiency. We found that accretion modified the structure of the companion: its radius and luminosity increased slightly, while the effective temperature decreased, without triggering significant expansion. We showed how average accretion rates grow with \( q \) but decline with increasing orbital period, consistent with lower wind densities at larger separations. The moderate thermal response is in line with expectations for low accretion rates \citep{2025ApJ...986..188M}, whereas higher rates are known to induce substantial expansion \citep{2025ApJ...986..188M, 2026NewA..12202475M}. The primary star experienced a luminosity decline due to envelope expansion and internal energy redistribution, underscoring the importance of accurate mass-loss prescriptions in massive binary models. We also derived an analytical relation linking the average accretion rate to orbital period and mass ratio, revealing a strong power-law dependence. A transition occured at \( P \lesssim 355 \)~days, where Roche-lobe overflow (RLOF) started to contribute significantly to the mass transfer. This threshold agreed with previous work \citep{2024ApJ...966..103P} and marks the shift from wind-dominated accretion in wide systems to hybrid RLOF--wind accretion in closer binaries, consistent with observational trends \citep{2024ApJ...966..103P}.

In this study, we extend our investigation of wind accretion in massive binary systems with evolved primarys undergoing intense mass-loss episodes. We consider primary stars with masses of $60$--$90\,\rm M_\odot$ paired with a $30\,\rm M_\odot$ companion, in orbits with periods of $455$--$1155$~days and eccentricities in the range $e = 0$--$0.6$. The models adopt eruptive mass-loss rates of $\dot{M}_{\rm w} = 10^{-2}$ and $10^{-1}\,\rm M_\odot\,\rm{yr}^{-1}$ over a duration of $1.5$~years, assuming purely wind accretion. From this grid, we derive the dependence of the accretion rate on mass ratio, mass-loss rate, orbital period, and eccentricity, and we present an analytical scaling relation calibrated to the simulation results. We further examine the thermal response of the companion star to the accreted material. Additionally, we explore scenarios in which the companion itself is undergoing stellar wind mass loss, allowing us to investigate the interplay between the winds of both binary components and its impact on the accretion dynamics.

\section{The physical ingredients of numerical Simulation}
\label{sec2}
We model 5 binary systems with orbital periods ranging from $P = 455$ to $1155$~days. The companion star is fixed at a mass of $M_{1} = 30\,\mathrm{M}_{\odot}$, while the primary mass is varied uniformly between $M_{2} = 60$ and $100\,\mathrm{M}_{\odot}$. The adopted physical parameters are inspired by the well-studied $\eta$~Carinae system \citep[e.g.,][]{2007MNRAS.378.1609K, 2008NewA...13..569K, 2016ApJ...825..105K}, which shows the wind accretion during the peristron passage. The primary star of $\eta$ Carinae undergoes extreme eruptive mass loss, reaching rates of order $\sim 1\,\rm M_\odot\,{\rm yr^{-1}}$ \citep{2010ApJ...723..602K}, while the binary has a long orbital period of $\simeq 5.53$ yr, implying a wide and detached configuration. Even during the giant eruption, the stellar radius is expected to remain well below its Roche lobe, so the system does not enter a RLOF phase.
Instead, the eruption drives a dense, slow outflow that includes the companion, which can accrete material directly from the wind. In such wide, non-contact binaries with extreme mass-loss rates, mass transfer is therefore expected to proceed via wind accretion rather than RLOF \citep[e.g.,][]{2001MNRAS.325..584S, 2007ASPC..372..397M, 2010ApJ...723..602K}.
Even if RLOF is considered for the $\eta$~Carinae system, the BHL and RLOF give approximately the same amount of accreted mass at periastron passage, though for BHL the accretion rate is smaller and lasts longer \cite{2009NewA...14...11K}.

Motivated by these observations, we adopt eruptive mass-loss rates of $\dot{M}_{\rm w} = 10^{-2}$-$10^{-1}\,\mathrm{M}_{\odot}\,\mathrm{yr^{-1}}$ for the primary. These values are consistent with the high mass-loss event in case $\eta$~Carinae outside of its Giant Eruption, during which even higher rates were achieved. In our earlier work, we explored lower mass-loss rates on the order of $10^{-3}\,\mathrm{M}_{\odot}\,\mathrm{yr^{-1}}$; here, we deliberately extend to higher values to probe their impact on stellar structure and accretion efficiency.\footnote{Throughout the paper, we denote by $\dot{M}_{\rm w}$ the artificial mass-loss rate imposed on the primary star. The companion's wind mass-loss rate is
denoted by $\dot{M}_{\rm comp,w}$ and is computed using the Dutch mass-loss prescription during the accretion phase.}
As shown by \citet{2024ApJ...974..124M, 2026ApJ...997...66M}, very high mass-loss rates (e.g., $\dot{M}_{\rm w} \sim 0.15\,\mathrm{M}_{\odot}\,\mathrm{yr^{-1}}$) can significantly cause structural changes in the primary star, which in turn influence wind accretion onto the companion. To capture this scenario, we select $\dot{M}_{\rm w} = 10^{-2}$--$10^{-1}\,\mathrm{M_{\odot}}\,\mathrm{yr^{-1}}$. In massive binaries, post-main-sequence evolution is often dominated by Case~A mass transfer for high-mass primarys \citep[e.g.,][]{2015A&A...580A..20S, 2015A&A...573A..71K, 2023A&A...672A.198S}. Since our goal is to isolate the physics of wind accretion, we intentionally construct binaries that avoid RLOF before the eruptive phase. Early Case~A mass transfer would substantially alter the structure and evolutionary state of the accretor, for example, through rejuvenation and mass accretion, thereby modifying its lifetime and internal profile when the primary reaches the eruptive mass-loss stage \citep{2023ApJ...942L..32R}. Such effects would introduce additional evolutionary channels that are not directly related to wind accretion and would complicate the interpretation of the accretion process itself. To prevent this, we first evolve both stars independently as single stars up to the post-main-sequence stage (listed in Table~\ref{T1}), and only then couple them using the \textsc{binary} module of \textsc{mesa}. By construction, the chosen orbital periods lie outside the RLOF boundary: in test simulations, we find that systems with initial periods less than 455 days undergo RLOF, while our model grid starts at longer periods ($\approx 455$) where the stars remain detached. This ensures that no Case~A mass transfer occurs and that the subsequent interaction is dominated by wind accretion.

 The simulation timescale is set to 1.5~years, which covers the critical orbital phase around periastron passage, where the accretion rate is expected to reach its maximum. The adopted primary and secondary masses ($60$-$100\,\mathrm{M}_{\odot}$ and $30\,\mathrm{M}_{\odot}$, respectively) are consistent with current estimates for the $\eta$~Carinae system \citep{2019MNRAS.486..926K}. We consider eccentric orbits with $e = 0.0$, 0.2, 0.4, and 0.6. Systems with higher eccentricities ($e > 0.6$) are excluded because the periastron separation becomes sufficiently small that the primary can approach or exceed its Roche lobe \citep[e.g.,][]{1983ApJ...268..368E}. Such configurations would lead to a hybrid interaction regime where wind accretion and RLOF overlap. Since we aim to isolate pure wind accretion, we restrict our study to eccentricities where the binary remains detached throughout the orbit.

\begin{table*}[t]
\centering
\caption{\textbf{Model grid and binary parameters adopted for wind accretion modeling.}}
\label{T2}
\begin{tabular}{|c|c|c|c|}
\hline
\hline
Binary system ($M_1$–$M_2$) & Orbital configuration & Primary mass-loss rate ($\rm M_\odot\,\mathrm{yr}^{-1}$) & Section \& Table \\
\hline

\multirow{2}{*}{$60$–$30$} 
& period (days): 455–1155 (8 runs) & $10^{-2}$ & \ref{3.3} \& \ref{T5} \\
& 455–1155 (8 runs) & $10^{-1}$ &  \ref{3.3} \& \ref{T5}  \\
\hline\hline

\multirow{2}{*}{$70$–$30$} 
& 455–1155 (8 runs) & $10^{-2}$ & \ref{3.3} \& \ref{T5} \\
& 455–1155 (8 runs) & $10^{-1}$ & \ref{3.3} \& \ref{T5} \\
\hline\hline

\multirow{2}{*}{$80$–$30$} 
& 455–1155 (8 runs) & $10^{-2}$ & \ref{3.3} \& \ref{T5} \\
& 455–1155 (8 runs) & $10^{-1}$ & \ref{3.3} \& \ref{T5} \\
\hline\hline

\multirow{2}{*}{$90$–$30$} 
& 455–1155 (8 runs) & $10^{-2}$ & \ref{3.3} \& \ref{T5} \\
& 455–1155 (8 runs) & $10^{-1}$ & \ref{3.3} \& \ref{T5} \\
\hline\hline

\multirow{2}{*}{$100$–$30$} 
& 455–1155 (8 runs) & $10^{-2}$ & \ref{3.1} \& \ref{T3} \\
& 455–1155 (8 runs) & $10^{-1}$ & \ref{3.1} \& \ref{T3} \\
\hline\hline

\multirow{2}{*}{$100$–$30$ }
& eccentricity: 0, 0.2, 0.4, 0.6 (4 runs) & $10^{-2}$ & \ref{3.2} \& \ref{T4} \\
& — & — & \\
\hline\hline

\multirow{2}{*}{$100$–$30$ ($P=555$ days)}
& companion wind included & $10^{-2}$ & \ref{3.4} \& - \\
& 455–1155 (8 runs), \& e=0&- & \\
\hline

\hline
\multicolumn{4}{|c|}{\textbf{Binary stellar parameters}} \\
\hline
Symbol & Parameter & Value & Units \\
\hline
$P$ & Period & 455–1155 & days \\
$e$ & Orbital eccentricity & 0, 0.2, 0.4, 0.6 & — \\
$\dot{M}_\mathrm{w}$ & Mass-loss rate & $10^{-2}$, $10^{-1}$ & $\rm{M_\odot}\,\mathrm{yr}^{-1}$ \\
$\alpha$ & Wind accretion efficiency & 1.5 & — \\
$\beta$ & Wind velocity exponent & 1.25 & — \\
$\eta_{\mathrm{max}}$ & Max transfer efficiency & 0.5 & — \\
$f_{\mathrm{pextra}}$ & Pextra factor & 1.5 & — \\
$q$ & Mass ratio ($q=M_2/M_1$) & 0.33, 0.37, 0.42, 0.5 & — \\
\hline
\hline
\end{tabular}
\end{table*}

To avoid RLOF during the early stages, we first evolve the primary and companion stars individually as single stars up to the onset of the post-main-sequence (post-MS) phase. At this point, the stellar parameters of both components are summarized in Table~\ref{T1}, and the corresponding evolutionary tracks are shown in \citet{Mukhija_2025} figure 1. The outputs from this stage are then used as inputs for the subsequent binary evolution calculations. Both stars are evolved to the same age to ensure a consistent binary configuration that naturally incorporates the relevant physical effects of binarity. During the single-star evolution phase, we adopt a low initial rotation rate of $0.1\,\Omega_{\rm crit}$ at solar metallicity ($Z = 0.0142$; \citealt{2009ARA&A..47..481A}). This setup results in a binary where the primary acts as the cooler component, while the companion is the hotter, more compact star. After inserting the single-star outputs into the binary module, we initiate the wind accretion phase and evolve the binary for $t = 1.5$ years. The binary system parameters relevant for wind accretion are summarised in Table~\ref{T2}. In these binary simulations, the primary loses mass at a constant, artificially imposed rate of $\dot{M}_{\rm w} = 10^{-2}$-$10^{-1}\,\mathrm{M}_{\odot}\,\mathrm{yr}^{-1}$, implemented directly within \textsc{mesa}. No additional wind mass-loss prescriptions are applied to either star during this phase, as the primary objective is to study accretion onto the companion purely due to the primary's imposed outflow. Including the companion's intrinsic wind would introduce further complexities, since the hotter secondary is expected to launch a stronger wind that could oppose the inflow and reduce the accretion efficiency, potentially driving the system into the sub-BHL regime \citep{2022MNRAS.516.3193K}. For this reason, the companion wind is initially neglected. In a later set of models, however, we also explore a scenario in which the secondary undergoes wind mass loss following the Dutch prescription simultaneously over the same 1.5 years interval of wind accretion. To avoid additional complications from rotation-induced winds, we use non-rotating stellar models during the binary evolution phase \citep[e.g.,][]{2000A&A...361..159M, 2012ARA&A..50..107L, 2014A&A...564A..57M}.

We employ OPAL Type~II opacity tables \citep{1996ApJ...464..943I}, which account for time-dependent changes in metal abundance, and apply an additional wind hook to model the imposed outflows. Convection is treated using the time-dependent convection (\textsc{tdc}) formalism with a fixed mixing-length parameter $\alpha_{\mathrm{MLT}} = 1.5$ \citep[e.g.,][]{2018ApJS..234...34P, 2023ApJS..265...15J}. This approach is also applied to sub-surface convective layers near opacity bumps. We activate the \textsc{mlt++} prescription to reduce the superadiabatic gradient in radiation-dominated regions, improving the stability and realism of the models. Convective boundaries are determined using the Ledoux criterion, which accounts for composition gradients, and semi-convection is included as a diffusive process \citep{1983A&A...126..207L} with a fixed efficiency parameter $\alpha_{\rm sc} = 1$ \citep{2006A&A...460..199Y, 2019A&A...625A.132S}. Overshooting is treated using an exponential formalism \citep{2000A&A...360..952H}, applied above convective cores to prevent numerical instabilities. We adopt $f_{0}(1) = 0.01$ and $f_{1}(1) = 0.345$ for the overshooting parameters.

\section{Results}
\label{sec3}

We evolve five binary systems using \textsc{mesa} to study wind accretion in massive binaries. The initial stellar parameters at the onset of the wind-accretion phase for both the primary and companion stars are listed in Table~\ref{T1}, while the relevant binary parameters are summarized in Table~\ref{T2}. Each simulation is evolved over a duration of $t = 1.5$~years, with orbital periods spanning $P = 455$-$1155$~days. For each mass ratio $q = M_{2}/M_{1}$ at ZAMS, we construct a grid of models by varying the orbital period across this range, resulting in five distinct systems for each $q$ value. A total of five mass ratios are considered, allowing us to systematically investigate the mechanics of wind accretion, its effect on both the primary and secondary stars, and its influence on key binary parameters such as orbital period and mass ratio.
To examine the accretion process in detail, we select a representative system with a $30\,\mathrm{M}_{\odot}$ companion and a primary mass in the range $M_{1}=60$-$100\,\mathrm{M}_{\odot}$, and an orbital period of $P = 555$~days. The parameter space is explored through four main cases; \textit{Case 1: Effect of mass-loss rate in circular orbits.}  \textit{Case 2: Effect of eccentricity.}  \textit{Case 3: Dependence on stellar mass and analytical scaling.} \textit{Case 4: Effect of companion wind}. This framework enables us to examine how wind accretion efficiency depends on mass-loss rate, orbital eccentricity, stellar mass, and the presence of a companion wind, providing a comprehensive view of wind-fed mass transfer in massive binary systems.

\begin{table*}[htbp]
\centering
\caption{Accreted mass $\Delta M$ and mass accretion rate $\dot{M}_{\rm acc}$ as a function of orbital period for the $100$-$30~\rm M_\odot$ system at $e=0$. Two mass loss cases are shown: $10^{-1}$ and $10^{-2}~\rm M_\odot~\mathrm{yr}^{-1}$. For the $\dot{M}_{\rm w}=10^{-1} \rm M_\odot,\mathrm{yr^{-1}}$ case, we also report the changes in $\log L$ and $\log T_{\mathrm{eff}}$ for both the primary and companion. Positive signs indicate an increase, while negative signs denote a decrease in the corresponding stellar parameters. For the visualization please see Figure~\ref{fig_2}.}
\label{T3}
\renewcommand{\arraystretch}{1.15}
\small
\begin{tabular}{|c|cc|cc|cc|cc|}
\hline\hline
\multirow{2}{*}{\textbf{Period (days)}} &
\multicolumn{2}{c|}{\textbf{$\Delta M$ ($10^{-6}~\mathrm M_\odot$)}} &
\multicolumn{2}{c|}{\textbf{$\dot{M}_{\rm acc}$ $(10^{-6}~\mathrm M_\odot~\mathrm{yr}^{-1})$}} &
\multicolumn{2}{c|}{{Primary ($\dot{M}_{\rm w}=10^{-1}$)}} &
\multicolumn{2}{c}{{Companion ($\dot{M}_{\rm w}=10^{-1}$)}} \\
\cline{2-9}
 & \textbf{$10^{-1}$} & \textbf{$10^{-2}$} & \textbf{$10^{-1}$} & \textbf{$10^{-2}$} &
 ${\Delta\log L}$ & ${\Delta\log T_{\mathrm{eff}}}$ &
 ${\Delta\log L}$ & ${\Delta\log T_{\mathrm{eff}}}$ \\
\hline
455  & 3.63 & 3.33 & 2.42 & 2.22 & $-0.4200$ & $+0.2600$ & $+0.0017$ & $-0.0100$ \\
555  & 2.82 & 2.57 & 1.88 & 1.71 & $-0.2645$ & $+0.2900$ & $+0.0002$ & $-0.0100$ \\
655  & 2.28 & 2.07 & 1.52 & 1.38 & $-0.2365$ & $+0.3148$ & $+0.0003$ & $-0.0018$ \\
755  & 1.90 & 1.72 & 1.26 & 1.15 & $-0.3355$ & $+0.2770$ & $+0.0003$ & $-0.0009$ \\
855  & 1.62 & 1.46 & 1.08 & 0.97 & $-0.3124$ & $+0.2833$ & $+0.0003$ & $-0.0009$ \\
955  & 1.40 & 1.26 & 0.93 & 0.84 & $-0.2966$ & $+0.2849$ & $+0.0003$ & $-0.0009$ \\
1055 & 1.23 & 0.98 & 0.82 & 0.65 & $-0.2215$ & $+0.3007$ & $+0.0004$ & $-0.0006$ \\
1155 & 1.09 & 0.98 & 0.72 & 0.65 & $-0.2215$ & $+0.3007$ & $+0.0004$ & $-0.0006$ \\
\hline\hline
\end{tabular}
\end{table*}

\subsection{Case 1; Effect of mass-loss rate in circular orbits}
\label{3.1}
\begin{figure*}
   \includegraphics[trim= 0.0cm 0.0cm 0cm 0.0cm,clip=true,width=1\textwidth]{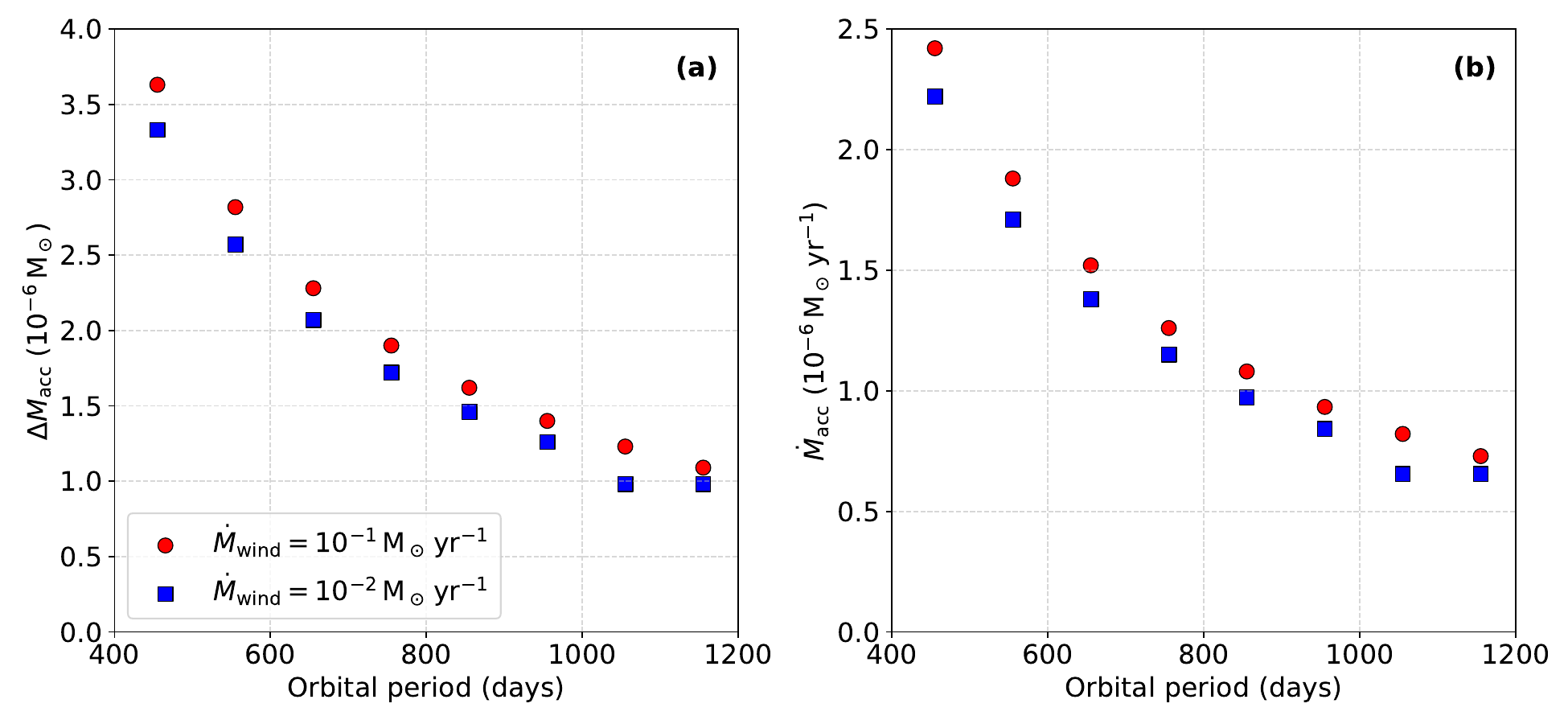} 
   \caption{Accreted mass $\Delta M$ (panel~a) and corresponding mass accretion rate $\dot{M}_{\rm acc}$ (panel~b) as a function of orbital period for the $100$--$30~\rm M_\odot$ binary system at $e=0$.}

   \label{fig_2}
\end{figure*}

\begin{table*}[htbp]
\centering
\caption{Mass accretion rates $\dot{M}_{\rm acc}$ for the $100$--$30~\rm M_\odot$ system with $\dot{M}_{\rm w} = $ $10^{-2}\,\rm M_\odot~\mathrm{yr}^{-1}$, for different orbital periods and eccentricities.}
\label{T4}

\renewcommand{\arraystretch}{1.15}
\begin{tabular}{|c|cccc|}
\hline
\hline
\multirow{2}{*}{\textbf{Orbital Period (days)}} & \multicolumn{4}{c}{$\dot{M}_{\rm acc}$ ($\rm 10^{-6}~ M_\odot\,\mathrm{yr}^{-1}$)} \\
\cline{2-5}
 & $e = 0.0$ & $e = 0.2$ & $e = 0.4$ & $e = 0.6$ \\
\hline
455  & $2.22 $ & $2.47 $ & $2.64 $ & $3.03 $ \\
555  & $1.71$ & $1.92 $ & $2.05 $ & $2.35$ \\
655  & $1.38 $ & $1.55 $ & $1.66 $ & $1.90$ \\
755  & $1.15 $ & $1.29 $ & $1.38 $ & $1.58$ \\
855  & $0.97$ & $1.1$ & $1.18 $ & $1.35$ \\
955  & $0.84 $ & $0.95 $ & $1.02 $ & $1.17$ \\
1055 & $0.65 $ & $0.83 $ & $0.89 $ & $1.03$ \\
1155 & $0.65$ & $0.74$ & $0.79$ & $0.91 $ \\
\hline
\hline
\end{tabular}
\end{table*}

In this case, we examined in detail a $100$-$30~\rm M_\odot$ binary system under two imposed mass-loss rates, $\dot{M}_{\rm w} = 10^{-2}$ and $10^{-1}~\rm {M_\odot} ~\mathrm{yr^{-1}}$, while varying the orbital period between 455 and 1155~days in circular orbits. This setup allows us to investigate how different eruptive mass-loss episodes influence the resulting accretion rate as a function of orbital period.
Figure~\ref{fig_2} illustrates the variation of the total accreted mass ($\Delta M$; panel a) and the mass accretion rate ($\dot{M}_{\rm acc}$; panel b) as a function of orbital period for this case. In both imposed mass loss episodes, $\Delta M$ decreases systematically with increasing orbital period. This is expected, since wider systems have lower wind densities, reducing the amount of material that can be captured by the companion star. For the shortest period system (455~days), $\Delta M$ is $\sim 3.6 \times 10^{-6}~\rm M_\odot$ for the $10^{-1}~\rm M_\odot~\mathrm{yr^{-1}}$ mass-loss case and $\sim 3.3 \times 10^{-6}~\rm M_\odot$ for the $10^{-2}~\rm M_\odot~\mathrm{yr^{-1}}$ case. By an orbital period of 1155~days, these values decrease to $\sim 1.1 \times 10^{-6}~\rm M_\odot$ and $\sim 9.8 \times 10^{-7}~\rm M_\odot$, respectively.
Similarly, the mass accretion rate $\dot{M}_{\rm acc}$ decreases with increasing orbital period. For short orbital periods, the accretion rate is $\sim 2.4 \times 10^{-6}~\rm M_\odot~\mathrm{yr}^{-1}$ for the $10^{-1}~\rm M_\odot~\mathrm{yr^{-1}}$ case and $\sim 2.2 \times 10^{-6}~\rm M_\odot~\mathrm{yr}^{-1}$ for the $10^{-2}~\rm M_\odot~\mathrm{yr^{-1}}$ case. At longer orbital periods (1155~days), the accretion rates drop to $\sim 7.3 \times 10^{-7}~\rm M_\odot~\mathrm{yr}^{-1}$ and $\sim 6.6 \times 10^{-7}~\rm M_\odot~\mathrm{yr}^{-1}$, respectively. 
Thus, varying the mass-loss from lower to higher value, i.e., $10^{-2}$ to $10^{-1}~\rm M_\odot~\mathrm{yr}^{-1}$, leads to a systematic increase in both the total accreted mass and the accretion rate at all orbital periods. The relative enhancement is most pronounced at shorter periods as shown in Figure~\ref{fig_2}, where the companion can efficiently intercept the denser wind, while at longer periods, both rates are reduced due to the geometric dilution of the wind. Table \ref{T3} presents the accreted mass and accretion rates corresponding to all orbital periods for both considered mass-loss rates. 
As shown in Figure~\ref{fig_2}, the relatively small change in the total accreted mass despite varying the primary mass-loss rate by an order of magnitude arises because wind accretion is also controlled by the transfer efficiency, rather than by $\dot{M}_{\rm w}$ alone. In BHL accretion: $\dot{M}_{\rm acc} \simeq \eta\,\dot{M}_{\rm w} $, where $\eta_{\max}$ is transfer efficiency . In our models, increasing $\dot{M}_{\rm w}$ also increases the wind density and typically the relative wind speed, which lowers $\eta$. This partial compensation causes $\dot{M}_{\rm acc}$, and hence the integrated accreted mass, to vary only weakly with $\dot{M}_{\rm w}$. In addition, the accretion prescription includes an explicit cap on the transfer efficiency (we adopt $\eta_{\max}=0.5$), further limiting the sensitivity of the accreted mass to $\dot{M}_{\rm w}$.
For a mass-loss rate of $\dot{M}_{\rm w} = 10^{-1}~\rm M_{\odot}~\mathrm{yr^{-1}}$, the primary star shows a substantial luminosity decrease of about 35-40\% and a corresponding increase in effective temperature of roughly 70-80\% across the examined orbital periods. The extent of these changes weakens slightly with increasing orbital period, indicating that systems with wider separations experience a smaller luminosity drop and a more moderate temperature rise. In contrast, the companion star exhibits only marginal variations, with its luminosity increasing by $\lesssim 0.1$\% and effective temperature by approximately $\simeq2$--$3$\%. These minimal changes suggest that even under strong mass-loss conditions, the companion remains stable, with only minor heating effects likely due to weak wind accretion. Table~\ref{T3} presents the variations in luminosity and effective temperature for the $\dot{M}_{\rm w} = 10^{-1}~\rm M_{\odot}~\mathrm{yr^{-1}}$ case during the wind-accretion phase, showing how these quantities change for both the primary and companion stars across different orbital periods.

For this case, we compare the accretion rates in Tables~\ref{T3} and \ref{T6} gives $\dot{M}_{\rm acc}/\dot{M}_{\rm BHL} \approx 2.3$–$2.6$ over the period range $P=455$–$1155$ days for the circular orbits. This indicates that our calculated wind accretion rates are systematically a few times higher than the classical BHL prediction. The classical BHL prescription assumes a homogeneous, steady, and supersonic wind flowing past a point-mass accretor with negligible feedback on the flow. Under these idealized conditions, the accretion rate depends mainly on the relative velocity and sound speed of the gas, and as a result, the classical BHL estimate can significantly misestimate the true mass accretion rate. Hydrodynamic studies of wind mass transfer show that in regimes where the flow is slow and structured, the effective accretion efficiency can be substantially higher than BHL expectations. Such deviations have been found in simulations of massive binaries and supergiant X-ray binaries, where complex wind dynamics and orbital effects lead to accretion rates that exceed simple analytical predictions \citep[e.g.,][]{Edgar2004Review, Bozzo2016, ElMellah2017}. For more details, see the appendix section.

\begin{figure*}
    \centering
    \includegraphics[width=1\linewidth]{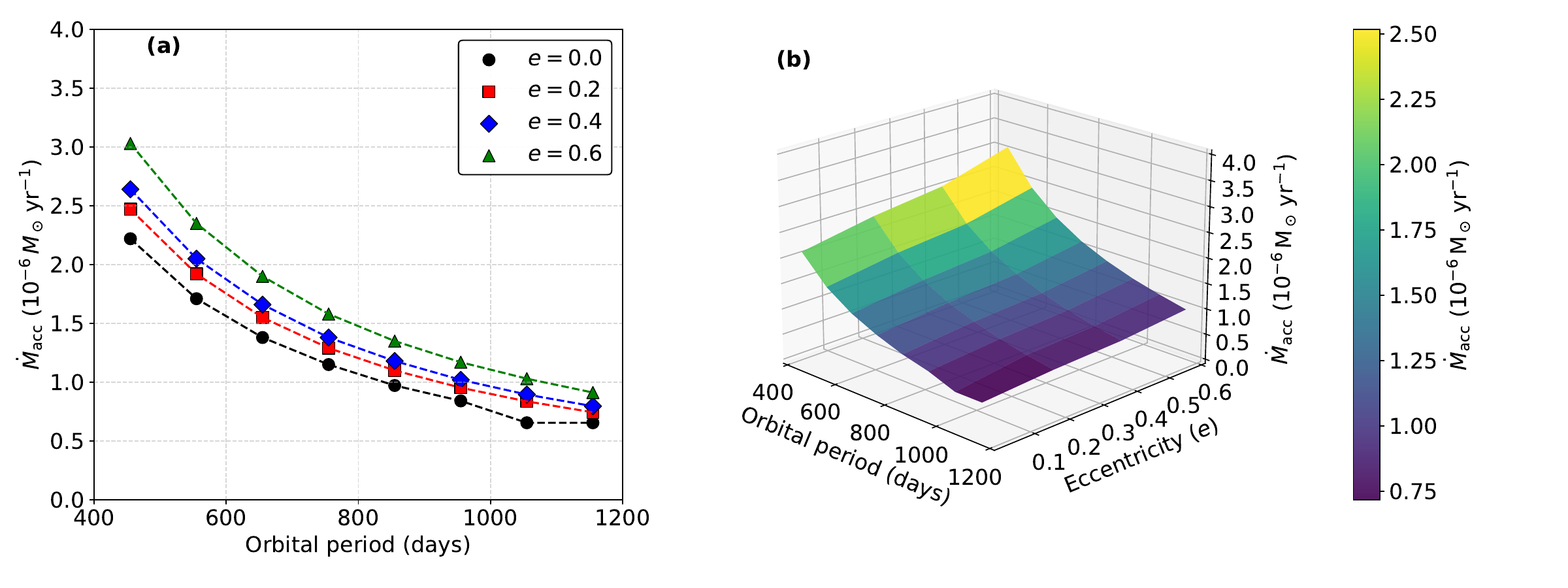}
    \caption{Here panel (a) shows the mass accretion rate, $\dot{M}_{\rm acc}$, as a function of orbital period for the $100$-$30~\rm M_\odot$ binary system, shown for different orbital eccentricities ($e = 0.0, 0.2, 0.4, 0.6$). The accretion rates are plotted in units of $10^{-6}\,\mathrm{M_\odot}\,\mathrm{yr}^{-1}$. Panel (b) shows the three-dimensional representation of $\dot{M}_{\rm acc}$ as a function of orbital period and eccentricity, illustrating the combined dependence of the wind-accretion efficiency on orbital separation and orbital eccentricity.}
    \label{fig_3}
\end{figure*}

\subsection{Case 2; Effect of eccentricity}
\label{3.2}

In this case, we investigate the effect of orbital eccentricity on wind accretion by adopting the same $100$-$30~\rm M_{\odot}$ binary system with a mass-loss rate of $\dot{M}_{\rm w} = 10^{-2}~\rm M_{\odot},\mathrm{yr^{-1}}$, while varying the eccentricity from $e = 0.0$ to $0.2$, $0.4$, and $0.6$. This approach enables us to quantify the influence of orbital geometry on accretion efficiency and overall mass-transfer dynamics. 

Figure~\ref{fig_3} presents the mass accretion rates $\dot{M}_{\rm acc}$ for the $100$--$30~\rm M_\odot$ system at $10^{-2}~\rm M_\odot~\mathrm{yr^{-1}}$, for different orbital periods and eccentricities. For a fixed orbital period, increasing the eccentricity leads to a systematic rise in the accretion rate. This trend reflects the fact that in eccentric systems, the companion spends part of its orbit at significantly smaller separations near periastron, where the local wind density and relative velocity favour more efficient wind accretion. For example, at a period of $455$~days, $\dot{M}_{\rm acc}$ increases from $2.22 \times 10^{-6}~\rm M_\odot~\mathrm{yr}^{-1}$ for $e=0$ to $3.03 \times 10^{-6}~\rm M_\odot~\mathrm{yr}^{-1}$ for $e=0.6$, corresponding to an enhancement of nearly $37\%$. A similar relative increase is observed across all orbital periods. At longer periods, the absolute accretion rates are lower overall due to wind dilution, but the effect of eccentricity remains evident, with higher eccentricities consistently yielding larger $\dot{M}_{\rm acc}$. This demonstrates that orbital eccentricity plays a significant role in enhancing mass accretion through wind focusing during periastron passages. The increase of $\dot{M}_{\rm acc}$ with eccentricity is not only because the companion gets closer at periastron, but because wind accretion depends very strongly on orbital separation. In the BHL mechanics, the accretion rate increases with wind density and decreases with relative velocity, and since the wind density rises rapidly at small radii, the accretion becomes strongly concentrated near periastron. Such an effect was shown for $\eta$~Carinae in \cite{2009NewA...14...11K}, where a mild change in eccentricity yielded a steep increase in the accretion rate and the total accreted mass. Even though the companion also spends time far away at apastron, the brief close approach dominates the orbit-averaged accretion, similar to how tidal forces are controlled by the periastron passage \citep{1981A&A....99..126H}. Table \ref{T4} shows the accretion rates for the $100$-$30~\rm M_\odot$ system for both mass-loss rate of $10^{-2}$ and $10^{-1}~\rm M_\odot~\mathrm{yr}^{-1}$, for different orbital periods and eccentricities.


\subsection{Case 3; Effect of massive systems}
\label{3.3}

In this case, we extend our grid to include five primary masses ($M_{1} = 100, 90, 80, 70,$ and $60~\rm M_{\odot}$) paired with a fixed $30~\rm M_{\odot}$ companion. Simulations are performed for two imposed mass-loss rates, $\dot{M}_{\rm w} = 10^{-2}$ and $10^{-1}~\rm M_{\odot}~\mathrm{yr^{-1}}$. From these models, we derive and calibrate an analytical relation that describes the average accretion rate as a function of the stellar and orbital parameters. In our previous work \citet{Mukhija_2025}, the accretion rate was expressed as a function of the orbital period and mass ratio only and given by power law
\begin{equation}
\dot{M}_{\rm acc}(t,P,q)
\approx 0.013~q^{1.1} \left(\frac{P}{\mathrm{day}}\right)^{-1.25}  \left(\frac{t}{\mathrm{yr}}\right)^{0.2},
\label{Eq1}
\end{equation}
where $q$ is the mass ratio, $P$ is the orbital period in days, and $t$ is the time period of wind accretion phase. We follow \citet{Mukhija_2025} in our simulations; the final mass ratio at the onset of wind accretion, denoted as \( q_f \), can be empirically related to the initial mass ratio at ZAMS, \( q \), by the following relation:
\begin{equation}
q_f \approx q^{0.60 \pm 0.03}.
\end{equation}
This power law approximation provides a practical conversion for applying our figures and fitting relations, originally expressed in terms of \( q \), to the actual mass ratio values at the start of the wind accretion. 
 
\begin{figure*}
  \includegraphics[trim= 0.0cm 0.0cm 0cm 0.0cm,clip=true,width=1\textwidth]{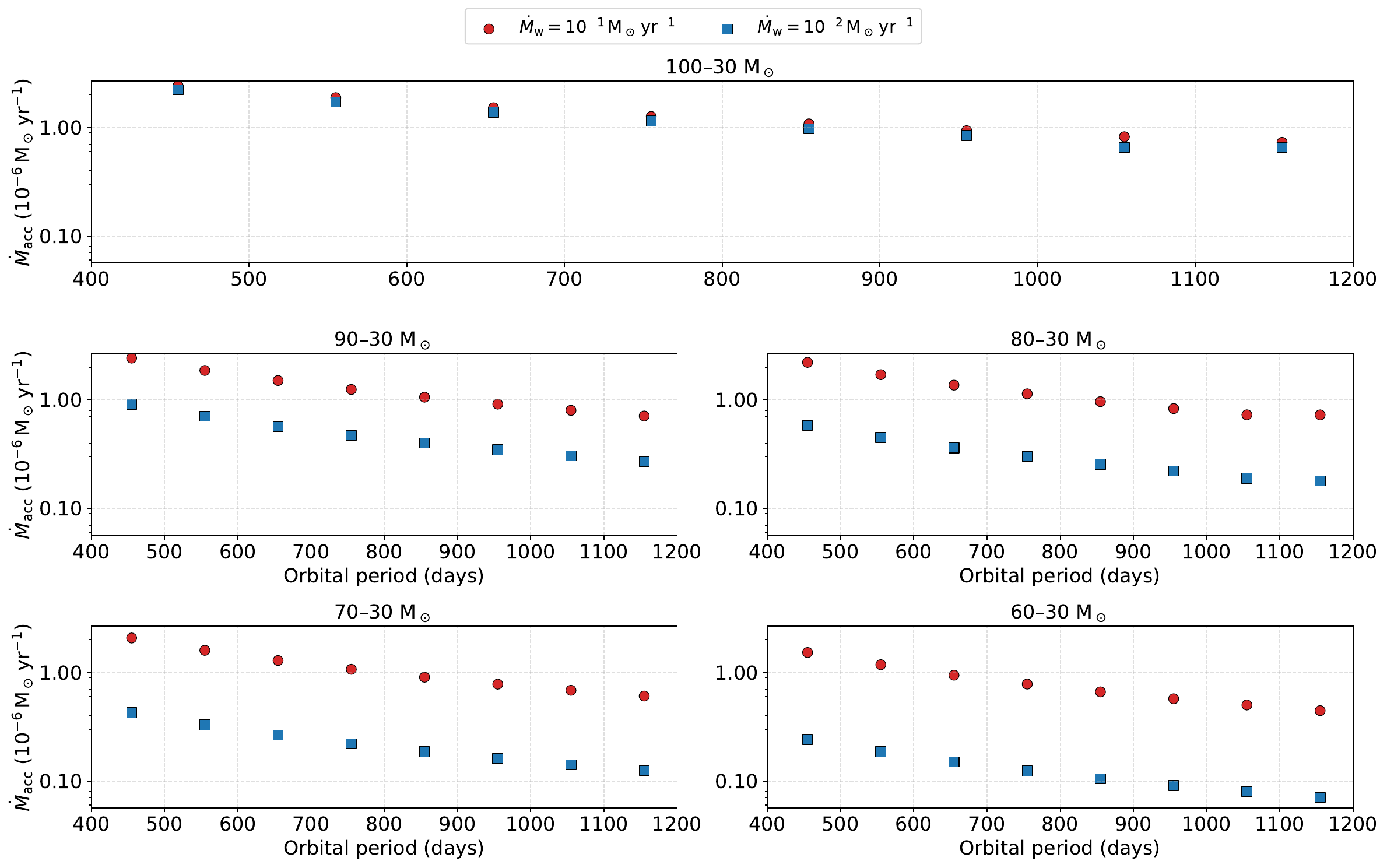} 
  \centering
  \caption{Mass accretion rates $\dot{M}_{\rm acc}$ as a function of orbital period for different binary systems at $e=0$. Each panel corresponds to a different primary mass. The red and blue points show the cases with mass loss rates of $10^{-1}$ and \ $10^{-2}~\rm M_\odot~\mathrm{yr}^{-1}$, respectively.}

\label{fig_4}
\end{figure*}

Here, we generalize this formulation to include dependencies on the mass ratio, orbital period, eccentricity, and mass-loss rate. In addition, we examine how the wind accretion efficiency varies with the primary star’s mass. Figure~\ref{fig_4} show the mass accretion rates $\dot{M}_{\rm acc}$ for different binary systems (100--30, 90--30, 80--30, 70--30, and 60--30~$\rm M_\odot$) at zero eccentricity, for two different mass loss rates ($10^{-1}$ and $10^{-2}~\rm M_\odot\,\mathrm{yr}^{-1}$). 
A clear trend emerges: for a given orbital period, the accretion rate is systematically higher in more massive systems. For example, at a period of 455~days, $\dot{M}_{\rm acc}$ reaches $2.42 \times 10^{-6}~\rm M_\odot\,\mathrm{yr}^{-1}$ for the 100---30~$\rm{M_\odot}$ system for the mass loss rate $10^{-1}~\rm M_\odot~\mathrm{yr}^{-1}$, while it drops to $1.53 \times 10^{-6}~\rm M_\odot\,\mathrm{yr}^{-1}$ for the 60--30~$\rm M_\odot$ system. This reflects the stronger gravitational focusing and higher wind densities in systems with more massive primaries, leading to more efficient accretion. Furthermore, the difference in accretion rate due to the two mass-loss rates ($10^{-1}$ and $10^{-2}~\rm M_\odot~yr^{-1}$) becomes increasingly significant for lower-mass systems. In the 60–30~$\rm M_\odot$ case, the relative difference in $\dot{M}_{\rm acc}$ between the two mass-loss scenarios is substantial across all periods. In contrast, for the 100–30~$\rm M\odot$ system, the two curves remain closer, indicating a partial saturation of the accretion efficiency in the high-mass regime. It might be possible that in these systems, the Bondi–Hoyle radius and gravitational focusing are primarily set by the relative velocity and orbital geometry rather than by the absolute wind density. Increasing the mass-loss rate, therefore, increases the available material but does not proportionally enlarge the effective accretion cross-section, leading to a weaker dependence of $\dot{M}_{\rm acc}$ on the total ejected mass. By comparison, lower-mass systems operate in a more regime, where the accretion efficiency responds more directly to the amount of material lost. In addition, the accretion efficiency of the companion may be moderated by its thermal response. If the accreted material approaches or exceeds the companion’s thermal adjustment timescale, the star cannot remain in thermal equilibrium, but according to our earlier study \citep{2026NewA..12202475M}, this scenario does not play a significant role here. Such effects could further contribute to the saturation regime in the high-mass systems. By comparison, lower-mass systems operate in a more supply-limited regime, where the accretion efficiency responds more directly to the amount of material lost.

\begin{center}
\begin{table*}[ht]
\caption{Accretion rates $\dot{M}_{\rm acc}$ as a function of orbital period for different systems (100--30, 90--30, 80--30, 70--30, and 60--30 $\rm M_\odot$) at $e=0$. Values are given for both mass loss rates $10^{-1}$ and  $10^{-2}$~$\rm M_\odot~\mathrm{yr}^{-1}$. }
\label{T5}
\small
\renewcommand{\arraystretch}{1.15}
\begin{tabular}{c cc cc cc cc cc}
\hline\hline
\multirow{2}{*}{\textbf{Period (days)}}
& \multicolumn{2}{c}{\textbf{100--30 $\rm M_\odot$}}
& \multicolumn{2}{c}{\textbf{90--30 $\rm M_\odot$}}
& \multicolumn{2}{c}{\textbf{80--30 $\rm M_\odot$}}
& \multicolumn{2}{c}{\textbf{70--30 $\rm M_\odot$}}
& \multicolumn{2}{c}{\textbf{60--30 $\rm M_\odot$}} \\
\cline{2-11}
& {10$^{-1}$} & {10$^{-2}$}
& {10$^{-1}$} & {10$^{-2}$}
& {10$^{-1}$} & {10$^{-2}$}
& {10$^{-1}$} & {10$^{-2}$}
& {10$^{-1}$} & {10$^{-2}$} \\
\hline
455  & 2.42 & 2.22 & 2.43 & 0.91 & 2.22 & 0.58 & 2.08 & 0.42 & 1.53 & 0.24 \\
555  & 1.88 & 1.71 & 1.87 & 0.70 & 1.71 & 0.45 & 1.60 & 0.32 & 1.18 & 0.18 \\
655  & 1.52 & 1.38& 1.51 & 0.56 & 1.37 & 0.36 & 1.29 & 0.26& 0.94 & 0.15 \\
755  & 1.26 & 1.15 & 1.25 & 0.47 & 1.14 & 0.30 & 1.07 & 0.22 & 0.78 & 0.12 \\
855  & 1.08 & 0.97 & 1.06 & 0.40 & 0.96 & 0.25 & 0.90 & 0.18 & 0.66 & 0.10 \\
955  & 0.93 & 0.84 & 0.91 & 0.34 & 0.83 & 0.22 & 0.78 & 0.16 & 0.57 & 0.09 \\
1055 & 0.82 & 0.65 & 0.80 & 0.30 & 0.73 & 0.19 & 0.68 & 0.14 & 0.50 & 0.07 \\
1155 & 0.72 & 0.65 & 0.71 & 0.27 & 0.72 & 0.05 & 0.60 & 0.12 & 0.44 & 0.07 \\
\hline\hline
\end{tabular}
\end{table*}
\end{center}

Finally, across all systems, $\dot{M}_{\rm acc}$ declines with increasing orbital period, as expected from the lower wind densities and weaker gravitational focusing at larger separations. The combined trends indicate that both the primary mass and the total ejected mass play important roles in determining the accretion rate, with the effect of the latter becoming more pronounced in lower-mass systems. Table~\ref{T5} show all the values of the accretion rate corresponding to the grid systems.

By considering and calibrating against the results presented in Tables~\ref{T4} and \ref{T5} (covering the period range $P=455$--$1155$~days, primary masses $M_1=60$--$100~\rm M_\odot$, $M_2=30~\rm M_\odot$, and eccentricities $e=0$--$0.6$), we adopt the following empirical, physics-informed scaling relation that reproduces the behavior of our simulation grid
\begin{equation}
\begin{split}
    \dot{M}_{\rm acc} &\approx A\,
    \left(\frac{M_1}{100\,\mathrm{M_\odot}}\right)^{1.71 \pm 0.17}
    \left(\frac{P}{555\,\mathrm{d}}\right)^{-1.3 \pm 0.07} \\
    &\times 
    \left(1 + (1.45 \pm 0.2)\, e\right)
    \left(\frac{\dot{M}_{\rm w}}{10^{-2}\,\mathrm{M_\odot\,yr^{-1}}}\right)^{0.242\pm 0.02} .
\end{split}
\label{Eq2}
\end{equation}
Here $M_1$ is the primary mass, $P$ the orbital period, and $e$ the eccentricity. With $A \simeq 1.3 \pm 0.06$ chosen so that the $100$--$30~\rm M_\odot$, $P{=}555$~d, $e{=}0$ case reproduces $\dot M_{\rm acc}\simeq 1.9\times10^{-6}\,\rm M_\odot\,\mathrm{yr}^{-1}$. Equation~\ref{Eq2} captures the main behaviours seen in the tables:
(i) \emph{Period:} $\dot M_{\rm acc}\propto P^{-1.3}$  matches the steep decline from $P{=}455$~d to $1155$~d in all systems.
(ii) \emph{Primary mass:} $\dot M_{\rm acc}\propto M_1^{1.71}$ accounts for the systematic boost from the $60$--$30$ to the $100$--$30\,\rm M_\odot$ systems at fixed $P$.
(iii) \emph{Eccentricity:} a linear factor $(1+1.45\,e)$ fits the modest $\sim$30--40\% enhancement between $e{=}0$ and $e{=}0.6$ at fixed $P$.
(iv) \emph{Wind mass-loss rate:} the accretion rate depends approximately linearly on the stellar wind mass-loss rate, $\dot M_{\rm acc}\propto \dot M_{\rm w}^{0.242}$, consistent with the expectation from Bondi-Hoyle-Lyttleton accretion. Across our two adopted mass loss rate, the resulting accretion rates scale nearly linearly with the corresponding wind mass-loss rates used in the simulations, confirming that the normalization of $\dot M_{\rm acc}$ is primarily set by $\dot M_{\rm w}$.

\subsection{Case 4; Effect of companion wind}
\label{3.4}

In this case, we explore a $100$-$30~\mathrm{M}{\odot}$ circular binary with $\dot{M}_{\rm w} = 10^{-2}~\mathrm{M}{\odot}~\mathrm{yr^{-1}}$ and include stellar wind mass loss ($\dot{M}_{\rm comp, w}$) from the companion during the accretion phase. In this setup, the primary’s artificial mass-loss rate remains fixed, and we add the companion wind following the Dutch prescription to assess how outflows from both stars interact and affect the accretion process. Since adding the stellar wind mass loss from the companion creates complexity, we did not include it in our earlier calculations. However, in this case, we recalculate the same $100$-$30~\mathrm{M}{\odot}$ system with a circular orbit and a mass-loss rate of $10^{-2}~\mathrm{M}{\odot}~\mathrm{yr^{-1}}$, now including the companion’s stellar wind using a Dutch scaling factor of 0.5. We find that adding a companion wind suppresses the accretion rate markedly across $P=455$--$755$~days. Relative to the no-companion wind baseline, the remaining accretion is $\simeq45.5\%$ at $P=455$~d (a $54.5\%$ decrease), $\simeq41.4\%$ at $555$~d ($58.6\%$ decrease), $\simeq25.4\%$ at $655$~d ($74.6\%$ decrease), and $\simeq13.9\%$ at $755$~d ($86.1\%$ decrease), yielding an average reduction efficiency of $\sim68\%$ over this period range. At longer periods ($P\gtrsim855$~d), the models return \emph{negative} accretion values, indicating that the accretion dynamics effectively shut down, consistent with material that might be captured being subsequently removed or prevented from settling by the companion’s wind, so that the net mass change of the accretor becomes negative. We emphasize that the magnitude of this suppression depends on the adopted Dutch wind scaling, with the factor set to $0.5$ here.

We also note that the behavior of accretion rate differs between short and long orbital periods, with two physical scenarios: one is the Hydrodynamic suppression of accretion effect, and the second is the net mass balance effect. At a small orbital period (455 to 755 days), the primary wind is dense and efficiently captured, so wind accretion dominates over the companion’s own mass loss, and the net accretion rate of the companion remains positive. In this regime, the companion wind reduces the accretion rate but does not prevent mass accretion; this regime has both effects. At larger orbital periods ($P\gtrsim 855$~d), however, the primary wind becomes less dense and hard to capture, reducing the intrinsic accretion efficiency. Companion wind can therefore eject more mass compared to accreted mass, thus leading to a negative net mass accretion rate. Both scenarios affect the companion star and are important for interpreting the results. A detailed separation of these effects requires fully three-dimensional modeling of wind–wind interaction physics, which is beyond the scope of the present study.

\section{Discussion}
\label{sec4}
Our study follows \citet{Mukhija_2025}, where we implemented BHL accretion in a massive binary using \textsc{MESA}. We assume the primary undergoes a high mass-loss episode ($10^{-3}~\rm M_{\odot}~yr^{-1}$) and the companion accreted part of the ejected material through wind capture. In our earlier work, we explored the dependence on mass ratio by fixing the primary mass, constructing a grid of companion masses, and varying the orbital period for a fixed primary mass-loss rate, restricting the analysis to circular orbits to isolate wind accretion dynamics. Here, we extend those studies in several directions. First, we consider two primary mass-loss rates to represent different eruptive episodes. Second,  we relax the circular-orbit assumption by introducing eccentric orbits. Third, we fix the companion mass and construct a grid of primary masses (thereby varying the mass ratio) to provide a complementary view relative to our previous setup and to give a more complete picture of parameter dependencies. Fourth, we include the companion’s own stellar wind mass loss. This expanded grid enables a broad set of numerical experiments that jointly assess how mass ratio, orbital period, eccentricity, and mass-loss rates shape wind-accretion efficiency and the resulting accretion dynamics in massive binaries.

Our numerical calculations of wind accretion in massive binaries are motivated by our earlier work. In \citet{2024ApJ...974..124M} we examined the consequences of a high mass-loss episode by prescribing an artificial outburst with $\dot{M}_{\rm w}=0.15\,\rm M_\odot\,\mathrm{yr^{-1}}$ from the outer envelope of a massive star. During the eruption the stellar luminosity decreased while the effective temperature increased, tracing a rapid excursion across the HR diagram and illustrating how severe mass removal reshapes the stellar structure. In the present study, applying an enhanced mass loss to the primary similarly leads to a luminosity drop and heating. We also previously investigated accretion onto a companion modeled as a single star, motivated by proposals that companion accretion in LBV binaries can power giant eruptions. In \citet{2025ApJ...986..188M} we computed a grid of accretors with masses $20$-$60\,\rm{M_\odot}$ subject to $\dot{M}_{\rm acc}=10^{-4}$-$10^{-1}\,\rm M_\odot\,\mathrm{yr^{-1}}$ over 20~years using \textsc{mesa}; for $\dot{M}_{\rm acc}\gtrsim10^{-2}\,\rm M_\odot\,\mathrm{yr^{-1}}$ the luminosity increased by roughly an order of magnitude, the envelope inflated, and the star moved to cooler $T_{\rm eff}$, whereas at lower rates the star remained comparatively hot with only modest brightening. Here, we compute the \emph{actual} wind-fed accretion rate in a binary and show that it depends sensitively on the stellar and orbital parameters (mass ratio, orbital period, eccentricity, and the primary’s mass-loss rate). For our configurations explored, the average wind-accretion rate is typically of order $\sim 10^{-6}\,\rm M_\odot\,\mathrm{yr^{-1}}$. At such rates, the accretor remains close to thermal equilibrium, and we do not find the large structural changes seen in our earlier high, $\dot{M}_{\rm acc}$ experiments, consistent with a wind-fed regime that is far less disruptive than the accretion scenarios considered previously in our work. 

Our results have direct implications for the evolution and observation of massive binaries in regimes where wind accretion dominates over Roche-lobe overflow. Such conditions are expected in wide \textsc{O+O} and \textsc{O+WR} binaries, luminous blue variable systems, and pre-interaction massive binaries undergoing enhanced wind mass loss \citep[e.g.,][]{2012Sci...337..444S, 2014ARA&A..52..487S}. In these environments, even modest changes in wind accretion efficiency can alter the mass ratio evolution, angular momentum budget, and rotational spin-up of the companion \citep[e.g.,][]{2012ARA&A..50..107L, 2013ApJ...764..166D, 2016A&A...588A..50M}. Our results demonstrate that wind accretion depends non-linearly on orbital separation and mass-loss rate. This has consequences for predicting the formation pathways of stripped stars, X-ray binaries, and gravitational-wave progenitors \citep[e.g.,][]{ElMellah2015, 2024ARA&A..62...21M}. Observationally, this regime is relevant to binaries where wind-wind interaction reshapes circumstellar structure, spectral diagnostics, and X-ray emission \citep[e.g.,][]{1992ApJ...386..265S, 2009MNRAS.396.1743P,2016A&A...590A.113G, 2023MNRAS.526.2167N,2024A&A...687A.197R, 2024A&A...687A.106B}. Systems with similar orbital periods but different wind mass-loss rates may therefore exhibit measurable differences in luminosity evolution, effective temperature shifts, and density structure, affecting line profiles and variability signatures. These effects are particularly important when interpreting wide interacting binaries such as $\eta$~Carinae, like systems and massive colliding-wind binaries, where wind accretion plays a significant role \citep{2009MNRAS.397.1426K}, highlighting the need for self-consistent wind accretion prescriptions to connect stellar evolution models with observable properties.

\section{Summary}
\label{sec5}
We model wind accretion in a grid of massive binary systems using \textsc{mesa}, examining how the accretion rate depends on both stellar and orbital parameters. We fix the companion mass at $M_2=30\,\rm M_\odot$ and vary the primary mass from $M_1=60$ to $100\,\rm M_\odot$, yielding mass ratios $q=M_2/M_1=\{0.30,\,0.33,\,0.37,\,0.42,\,0.50\}$. The orbital period spans $P=455$--$1155$~days; we avoid $P=355$~days because our earlier work indicates that such configurations may enter RLOF. We consider eccentricities $e=0.0,\,0.2,\,0.4,$ and $0.6$; higher values are not explored to prevent potential overlap between wind accretion and RLOF. For the primary, we impose two eruptive mass-loss rates, $\dot{M}_{\rm w} = 10^{-2}$ and $10^{-1}\,\rm M_\odot\,\mathrm{yr^{-1}}$, applied over a duration of $1.5$~years. Across this grid, the accretion rate onto the companion increases systematically with mass ratio, as expected from wind-accretion theory. The time-dependent evolution of the accretion rate during the $\sim$1.5~yr accretion phase is nearly constant and was presented in detail in our previous work \citet{2025arXiv250912725M}; we therefore refer the reader to that study for details. A more massive companion produces a deeper gravitational potential and stronger gravitational focusing, allowing it to capture a larger fraction of the primary’s wind and thus accrete more efficiently.

During Case 1, the $100$--$30\,\rm M_\odot$ circular binary, both the total accreted mass and the accretion rate decline systematically as the orbital period increases, reflecting geometric wind dilution and weaker gravitational focusing at wider separations. Raising the primary’s imposed mass-loss rate consistently enhances accretion at all periods, with the largest relative gains occurring at shorter periods where the companion samples denser, slower wind. The primary exhibits a strong structural response to the high outflow, whereas the companion remains largely stable, showing only minor changes consistent with modest wind capture and near-thermal equilibrium. Overall, accretion efficiency in this configuration is chiefly regulated by orbital separation and the primary’s mass-loss rate, and even during strong eruptive episodes, the companion’s internal structure is only weakly affected, as shown in Figure \ref{fig_2}. In case 2, introducing orbital eccentricity systematically (as shown in Figure~\ref{fig_3}) enhances wind accretion at any given period, owing to closer separations and denser, slower wind around periastron that increase capture efficiency. While absolute accretion rates still decline toward longer periods because of geometric dilution, the relative boost from eccentricity persists across the grid, indicating that periastron focusing is an amplifier of wind-fed mass transfer. We deliberately exclude very high eccentricities to avoid regimes where wind accretion overlaps with RLOF, allowing us to isolate the eccentricity-driven enhancement within the wind-accretion dynamics. For a fixed orbital period, moving from a circular orbit to an eccentric one consistently increases the wind-accretion rate. In broad terms, relative to $e=0$, the accretion rate is typically higher by about $10$-$15\%$ at $e=0.2$, about $20$-$25\%$ at $e=0.4$, and about $35$-$40\%$ at $e=0.6$. In case 3, as shown in Figure \ref{fig_4}, expanding the grid to multiple primary masses confirms a clear, coherent picture of wind-fed accretion in massive binaries. For a fixed companion mass, systems with more massive primaries consistently achieve higher accretion rates at the same orbital period, reflecting stronger gravitational focusing and denser winds. At the same time, accretion declines steeply with increasing period because wider separations dilute the wind and weaken focusing. Eccentricity provides a boost across the board via periastron passages, while the overall normalization of the accretion rate tracks the primary’s outflow almost linearly. The impact of changing the primary’s mass-loss rate is most pronounced in the lower-mass primary systems; in the higher-mass primaries, gravitational focusing dominates and the two mass-loss cases lie closer together. Taken together, these trends offer a practical, physically grounded guide: accretion is maximized in short-period, high-mass systems and is further enhanced by eccentricity; it weakens at long periods and becomes increasingly sensitive to the absolute wind supply in lower-mass systems.
While in case 4, including a companion wind in the $100$--$30\,\rm M_\odot$ circular system with $\dot{M}_{\rm w}=10^{-2}\,\rm M_\odot\,\mathrm{yr^{-1}}$ markedly surpres wind-fed accretion across $P=455$--$755$~days and effectively shuts it down at longer periods, where the net accretion becomes negative. This behaviour is consistent with the companion wind stripping or preventing the settling of material that would otherwise be captured. The magnitude of the suppression is set by the adopted wind scaling; weaker companion winds (smaller Dutch factors) would increase the accretion efficiency, partially restoring the no companion-wind baseline. In practice, accretion efficiency in this configuration improves when the companion wind is reduced and independently of the wind, when the orbit favours shorter periods or modest eccentricities that enhance periastron focusing and help the accretor overcome outflow opposition.

Another key outcome of our modeling is the primary’s structural response to strong mass loss. As the primary is stripped, its luminosity declines while the effective temperature rises, consistent with the photosphere receding into hotter layers and a redistribution of internal energy. This behavior underscores the need for realistic, time-dependent wind prescriptions in stellar-evolution calculations of massive binaries, especially when high outflow rates are present. In parallel, our calibrated scaling for the phase-averaged accretion rate, built across primary mass, orbital period, eccentricity, and primary wind strength, offers a practical way to estimate wind-accretion efficiency in diverse systems. The clear power-law-like sensitivities highlight that accretion is strongly dependent, reinforcing the importance of accounting for system, specific parameters (and, where relevant, a companion wind) when modeling mass transfer in massive binaries.

\section{Acknowledgments}

We thank an anonymous referee whose comments and
suggestions have helped to improve this paper.
We acknowledge the Ariel HPC Center at Ariel University for providing computing resources that have contributed to the research results reported in this paper. BM acknowledges support from the AGASS Center at Ariel University. The \textsc{mesa} inlists, and input files to reproduce our simulations and associated data products are available on Zenodo (10.5281/zenodo.18202724).

\section{appendix}

\subsection{Classical Bondi--Hoyle--Lyttleton estimate}
\label{sec:BHL}

 We compare our wind accretion rates to the classical Bondi--Hoyle--Lyttleton (BHL) prescription for accretion onto a point mass moving through an ambient medium \citep{1944MNRAS.104..273B}. In its standard form, the BHL accretion rate onto an accretor of mass $M_2$ (companion star) is
\begin{equation}
\dot{M}_{\rm BHL}
=
\frac{4\pi\,G^2\,M_2^2\,\rho}{\left(v_{\rm rel}^2+c_s^2\right)^{3/2}},
\label{eq:BHL_general}
\end{equation}
where $\rho$ is the gas density at the companion position, $v_{\rm rel}$ is the relative velocity between the companion and the incoming flow, $c_s$ is the sound speed, and $G$ is the gravitational constant. For hot-star winds, the flow is typically supersonic at the orbit ($c_s\ll v_{\rm rel}$), so Eq.~(\ref{eq:BHL_general}) reduces to
\begin{equation}
\dot{M}_{\rm BHL}
\simeq
\frac{4\pi\,G^2\,M_2^2\,\rho}{v_{\rm rel}^{3}}.
\label{eq:BHL_supersonic}
\end{equation}
Assuming the primary star wind is approximately spherical at the location of the companion, the wind density at orbital separation $a$ is
\begin{equation}
\rho_{\rm w(a)}=\frac{\dot{M}_{\rm w}}{4\pi a^2 v_{\rm w}},
\label{eq:rho_w}
\end{equation}
where $\dot{M}_{\rm w}$ is the mass-loss rate and $v_{\rm w}$ is the wind speed of  primary star. For a circular orbit ($e=0$), the separation follows Kepler's third law,
\begin{equation}
a(P)=\left[\frac{G\,(M_1+M_2)}{4\pi^2}\,P^2\right]^{1/3},
\label{eq:Kepler_a}
\end{equation}
where $P$ is the orbital period and $M_1$ and $M_2$ are the primary and companion masses, respectively. The orbital speed of the companion about the center of mass is
\begin{equation}
v_2=\frac{2\pi a}{P}\,\frac{M_1}{M_1+M_2}.
\label{eq:v2}
\end{equation}
The relevant relative speed between the wind flow and the companion is then approximated for the circular orbits by: 
\begin{equation}
v_{\rm rel}\simeq\left(v^2_{\rm w}+v_2^2\right)^{1/2},
\label{eq:vrel}
\end{equation}
where we have neglected the sound speed in the wind compared to the bulk flow speed. 
Substituting Eq.~(\ref{eq:rho_w}) into Eq.~(\ref{eq:BHL_supersonic}) yields a convenient BHL estimate,
\begin{equation}
\dot{M}_{\rm BHL}
\simeq
\frac{4\pi G^2 M_2^2}{v_{\rm rel}^3}\,
\frac{\dot{M}_{\rm w}}{4\pi a^2 v_{\rm w}}
=
\frac{G^2 M_2^2\,\dot{M}_{\rm w}}{a^2\,v_{\rm w}\,v_{\rm rel}^3}.
\label{eq:BHL_wind}
\end{equation}
Equation~(\ref{eq:BHL_wind}) makes clear the key scalings:
(i) $\dot{M}_{\rm BHL}\propto \dot{M}_{\rm w}$ (linear in the primary mass-loss rate),
(ii) $\dot{M}_{\rm BHL}\propto a^{-2}$ (declines with separation),
and (iii) $\dot{M}_{\rm BHL}\propto v_{\rm rel}^{-3}$ (strong sensitivity to the flow speed). Because our primary mass loss is imposed, the wind velocity profile is not solved self-consistently. For a reference BHL estimate, we adopt a characteristic wind speed based on the primary gravitational potential,
\begin{equation}
v^2_{\rm w} \simeq \frac{2\beta_{\rm w} G M_1}{R_1},
\label{eq:vw_beta}
\end{equation}
where $R_1$ is the primary star radius at the onset of wind accretion and $\beta_{\rm w}$ is represents the wind momentum flux parameter. With $\beta_{\rm w}=1.25$, $M_1=100\,\rm M_\odot$, and $R_1=100\,\rm R_\odot$, we obtain $v_{\rm w} \simeq 691~\mathrm{km\,s^{-1}}$. We evaluate Eqs.~(\ref{eq:Kepler_a})--(\ref{eq:BHL_wind}) for the $M_1=100\,\rm M_\odot$, $M_2=30\,M_\odot$ circular binary. For $\dot{M}_{\rm w}=10^{-2}\,\rm M_\odot\,\mathrm{yr^{-1}}$ and the adopted $v_{\rm w}=691~\mathrm{km\,s^{-1}}$, the BHL estimates are given in Table~\ref{T6}.

\begin{table}[t]
\centering
\caption{Classical BHL accretion rate for different orbital periods assuming $v_{\rm w} \simeq 691~\mathrm{km\,s^{-1}}$.}
\label{T6}
\begin{tabular}{cc}
\hline
\hline
Orbital period $P$ (days) & $\dot{M}_{\rm BHL}$ ($ \times10^{-7}\, \rm M_\odot\,\mathrm{yr^{-1}}$) \\
\hline
455  & $8.71$ \\
555  & $6.71$ \\
655  & $5.40$ \\
755  & $4.48$ \\
855  & $3.80$ \\
955  & $3.29$ \\
1055 & $2.88$ \\
1155 & $2.56$ \\
\hline
\end{tabular}
\end{table}

For the higher primary mass-loss case $\dot{M}_{\rm w}=10^{-1}\,\rm M_\odot\,\mathrm{yr^{-1}}$, Eq.~(\ref{eq:BHL_wind}) implies a simple linear scaling, so the values in Table~\ref{T6} increase by a factor of 10. In Table~\ref{T3} (the $\dot{M}_{\rm w}=10^{-2}\,\rm M_\odot\,\mathrm{yr^{-1}}$ column), our computed accretion rates span
$ \dot{M}_{\rm acc}\simeq (0.65\text{--}2.22)\times10^{-6}\ \rm M_\odot\,\mathrm{yr^{-1}}$ \rm for orbital period of P= 1155 - 455 days.

Comparing accretion rate in Tables~\ref{T3} \& \ref{T6} yields
\begin{equation}
\frac{\dot{M}_{\rm acc}}{\dot{M}_{\rm BHL}}\approx 2.3\text{--}2.6
\qquad \text{for } P=455\text{--}1155~\mathrm{d},
\label{eq:ratio_approx}
\end{equation}\\
i.e., our calculated wind accretion rates are a factor of a few larger than the simple BHL estimate under the adopted wind-speed normalization in Eq.~(\ref{eq:vw_beta}). We emphasize that this comparison is intended as a reference scaling: the classical BHL model assumes a steady, homogeneous upstream flow and neglects wind acceleration, non-uniform density or velocity structure, and detailed binary geometry.

\bibliography{Ref}{}
\bibliographystyle{aasjournal}
\end{document}